\documentclass[12pt]{article}
\usepackage{latexsym}
\usepackage{amssymb}
\usepackage{cite}
\usepackage{epsfig}
\title{\LARGE $Z^0$-Boson Decays in a
Strong Electromagnetic Field {\footnote{Published in:
Yad. Fiz. {\bf 72} (2009) 1078 , [Phys. At. Nucl. {\bf
72} (2009) 1034].}}}
\author{\large A.V.Kurilin{\footnote{E-mail address:
kurilin@mail.ru}}\\ Moscow Technological Institute \footnote{on leave from Moscow State Open Pedagogical University},\\ Leninsky Prospect, 38a, Moscow, 119334, Russia\\}

\begin{document}
\begin{titlepage}
\maketitle \thispagestyle{empty}

\begin{abstract}{The probability of $Z^0$-boson decay to
a pair of charged fermions in a strong electromagnetic
field, $Z^0\rightarrow \bar f f$, is calculated. On
the basis of a method that employs exact solutions to
relativistic wave equations for charged particles, an
analytic expression for the partial decay width
$\Gamma(\varkappa)=\Gamma(Z^{0}\rightarrow \bar f f)$
is obtained at an arbitrary value of the parameter
$\varkappa=e M_Z^{-3}\sqrt{-(F_{\mu\nu}q^\nu)^2}$,
which characterizes the external-field strength. The
total $Z^0$-boson decay width in an intense
electromagnetic field, $\Gamma_Z(\varkappa)$, is
calculated by summing these results over all known
generations of charged leptons and quarks. It is found
that, in the region of relatively weak fields
($\varkappa <0,06$), the field-induced corrections to
the standard $Z^0$-boson decay width in a vacuum do
not exceed 2\%. As $\varkappa$ increases, the total
decay width $\Gamma_Z(\varkappa)$ develops
oscillations against the background of its gradual
decrease to the absolute-minimum point. At
$\varkappa_{\rm min}=0,445$, the total $Z^0$-boson
decay width reaches the minimum value of
$\Gamma_Z(\varkappa_{\rm min})=2,164$ GeV, which is
smaller than the $Z^0$-boson decay width in a vacuum
by more than 10\%. In the region of superstrong fields
($\varkappa > 1$), $\Gamma_Z(\varkappa)$ grows
monotonically with increasing external-field strength.
In the region $\varkappa > 5$, the
$t$-quark-production process $Z^{0}\rightarrow \bar t
t$, which is forbidden in the absence of an external
field, begins contributing significantly to the total
decay width of the $Z^0$- boson.}
\end{abstract}
\end{titlepage}

\large
\section{INTRODUCTION}

It is not an exaggeration to say that $Z^0$-bosons
along with their charged partners, $W^{\pm}$-bosons,
have been at the focus of attention in high-energy
physics over more than the past two decades. The
properties of these particles, which are mediators of
weak interaction between leptons and quarks, were
studied in detail in experiments at the LEP and SLC
electron-positron colliders. In a series of
experiments performed between 1989 and 1995 at CERN
\cite{LEP1}, enormous statistics of observations of
$Z^0$-boson production and decays were collected. In
all, approximately $2\cdot 10^6$ events associated
with leptonic modes of $Z^0$-boson decay and more than
$1.5\cdot 10^7$ events of transformation of these
unstable particles into hadrons were detected. The
results of these investigations made it possible to
determine, among other things, the mass of the
$Z^0$-boson, $M_Z=91,1876\pm 0,0021\ \mbox{GeV}$, and
its total decay width, $\Gamma_Z=2,4952\pm 0,0023\
\mbox{GeV}$, to an extremely high degree of precision
\cite{PDG-2006}. Many other parameters of the process
$e^+e^-\rightarrow Z^0 \rightarrow \bar f f$ were also
analyzed both in the vicinity of the $Z$-resonance and
far from it \cite{Z-Resonance}. The precision of
experimental data reached in recent years has given a
motivation to theoretical physicists for performing
formidable work on calculating radiative corrections
to the processes involving $Z^0$-boson production and
decay (see, for example, \cite{UFN-96, Harlander98,
Precision, 2-loops-tt} and references therein).

Not only does $Z^0$-boson physics provide a precision
test for Standard Model predictions concerning
particle interactions, but it is also a unique tool
for seeking manifestations of new physics (beyond the
Standard Model), which will replace sooner or later
the generally accepted $SU(3)\times SU(2)\times U(1)$
scheme. Despite the impressive successes of the
Standard Model in adequately describing a formidable
set of experimental data, there is no doubt among the
physics community that the modern model of fundamental
interactions cannot be conclusive because of a number
of fundamental theoretical drawbacks. These include,
first, an enormous number (more than 19!) of
independent input parameters --- coupling constants,
the masses of quark and leptons, the parameters of the
Cabibbo-Kobayashi-Maskawa mixing matrix, etc; second,
the instability of the mass of the as-yet-undiscovered
Higgs boson with respect to quadratically divergent
radiative corrections and the allied hierarchy
problem; and, third, the isolated character of
gravitational interactions, which do not fit in the
existing quantum model. This is not the whole story,
however. Paradoxically as it is, the main drawback of
the Standard Model at the present time is that it is
in perfect agreement with all experimental data
accumulated thus far, providing no hint as to the
nature of new physics that will have to replace it.
This kind of "recess" in the experimental development
of physics theory caused a flash of creative activity
in theoretical physics in attempts at guessing vague
outlines of a new theory of everything. By no means do
Grand Unification theories, supersymmetry,
supergravity, superstrings, supermembranes, and
$M$-theory exhaust the list of ideas that require an
experimental verification. The questions that a future
theory should answer include that of the origin of
particle masses, that of why our space-time is
four-dimensional, that of why several generations of
leptons exist, that of how many flavors quarks have,
and that of how many neutrino flavors exist. Even
today, physicists try to find answers to many of these
questions by analyzing available experimental data.

For example, the total invisible width of
$Z^0$-bosons, $\Gamma_{\rm inv}=499,0\pm 1,5\
\mbox{MeV}$, with respect to the decays of these
particles through channels that are rather difficult
to detect plays an important role in $Z^0$-boson
physics. In the experiments performed at the LEP
accelerator, the invisible width $\Gamma_{\rm inv}$
was determined indirectly as the difference of the
total decay width of the $Z^0$-boson and its total
width with respect to all observed decays. In the
Standard Model, the invisible decay width of the
$Z^0$-boson is interpreted as the partial width with
respect to decays to neutrinos of various flavors: $$
\Gamma_{\rm inv}=\sum_{i}^{N_{\nu}}
\Gamma(Z^0\rightarrow\nu_i\tilde\nu_i)=N_{\nu}
\Gamma_{\nu\tilde \nu} $$

This quantity can be considered as some kind of a
counter of neutrino generations, and its precise
measurement is of particular interest for new physics
(see, for example, \cite{Zinv}). It is intriguing that
the present-day experimental value of the number of
neutrinos from the invisible decay width of the
$Z^0$-boson, $N_{\nu}=2.92\pm 0,06$ \cite{PDG-2006},
differs somewhat from the result expected in the
Standard Model, $N_{\nu}=3$.

Exotic channels of $Z^0$-boson decay frequently
provide unique tests for some fundamental principles
of modern physics. For example, analysis of the
forbidden (in a vacuum) $Z^0$-boson decays to photons
($Z^0\rightarrow\gamma\gamma$) and gluons
($Z^0\rightarrow g g$) makes it possible to establish
a quantitative measure of the principle of quantum
indistinguishability of integer-spin particles ---
Bose symmetry for photons and gluons \cite{Z_gamma}.
According to the Landau-Yang theorem
\cite{Landau-Yang}, a massive vector particle cannot
decay to massless vector states. In an external
electromagnetic field, however, the reactions
$Z^0\rightarrow\gamma\gamma$ and $Z^0\rightarrow g g$
become possible \cite{Tinsley02}.

The problem of searches for new possible
manifestations of supersymmetry is also tightly
related to $Z^0$-boson physics. There is still a hope
for finding relatively light superparticles produced
in rare channels of $Z^0$-boson decay. As an example,
one can consider the reaction $Z^0\rightarrow \tilde g
\tilde g$, in which a pair of light gluinos, $\tilde
g$, having a mass of about $m_{\tilde g}=12\div
16$~GeV appear. This process, which is extensively
discussed in the literature (see \cite{Z_gluinos} and
references therein), is of special interest in view of
prospects for observing new-physics manifestations at
next-generation $p\bar p$-colliders. Projects of
experiments with $Z^0$-bosons at the Large Hadron
Collider(LHC) also caused interest in studying the
properties of these particles in the presence of a
quark-gluon plasma, which is quite an unusual
environment of \cite{ZinPlasma}.

In view of the aforesaid, it is of particular interest
to discuss some other new phenomena in $Z^0$-boson
physics that manifest themselves under unusual
conditions. In this study, we investigate changes in
partial decay widths of the $Z^0$-boson in the
presence of strong electromagnetic fields. Interest in
such investigations is explained both by astrophysical
applications and by applications in the physics of
relativistic particles channeling through single
crystals (so-called channeling of particles
\cite{UFN-89}). It is well known that the strength of
electric fields generated by the axes and planes of
single crystals may reach formidable values of ($E\ge
10^{10}\;$V/m) \cite{Cern-94-05} over macroscopic
distances. At the same time, it is well known that
astrophysics studies exotic objects (neutron stars,
white dwarfs) that, at the latest stages of their
evolution, can undergo a strong compression, and this
leads to a significant increase in the magnetic field
strength inside them ($H\ge 10^8\div 10^{13}\;$ G
\cite{Shapiro}). At such enormous values of the
external field strength, the physics of quantum
processes changes significantly. External fields
frequently remove the forbiddance of specific
reactions whose occurrence is impossible in a vacuum.

The first investigations into the physics of
relativistic particles in the presence of a strong
external field were performed within quantum
electrodynamics (see, for example,
\cite{Sokolov-Ternov, Ritus-79,
Ternov-Khalilov-Rodionov}). Later, the technique of
calculations that was developed there was also used in
non-Abelian gauge theories \cite{UFN-97}. Respective
calculations rely on the method of exact solutions to
the wave equations for charged particles, which makes
it possible to take into account the interaction with
the electromagnetic field beyond standard perturbation
theory. In this case, the wave functions for all
charged particles and their propagators are modified
in quite a unwieldy manner, depending on the
configuration of an external field
\cite{Kurilin-1999}. Although the resulting
expressions are cumbersome and although the
calculations are quite involved, the ultimate results,
which are valid at arbitrarily high values of the
external-electromagnetic-field strength, carry
information that cannot be obtained by perturbative
methods, and this is an obvious advantage of the
method. From this point of view, investigation of
quantum processes in superstrong fields provides a
unique possibility of analyzing the self-consistency
of the physical theory globally in the case where an
expansion in a small parameter is impossible.

The method of calculations that is used below is based
on the crossed-field model, which was successfully
applied in the previous studies of the present author
to studying photino-pair production
\cite{Kurilin-Ternov} and to analyzing the decays of
$W$-bosons in an external field \cite{Kurilin-2004}.

\section{PROBABILITY FOR $Z^0$-BOSON DECAY\\ IN EXTERNAL FIELDS}

In the leading order of perturbation theory in the
coupling constants $g_V$ and $g_A$, the matrix element
of $Z^0$-boson decay to a pair of charged leptons,
$\ell^{\pm}$, is given by
\begin{equation}
\label{SZ-matrix} S_{fi}=i\int d^4 x
\overline{\Psi}_{\ell^-}(x,p)\gamma^\mu (g_{V}+\gamma^5
g_{A})\Psi_{\ell^+}^c(x,p') Z_\mu(x,q).
\end{equation}
The presence of an external electromagnetic field is
taken into account via choosing specific wave
functions $\Psi_{\ell^-}(x,p)$ and
$\Psi_{\ell^+}(x,p')$ for the charged leptons
$\ell^{\pm}$. This approach, which, in particle
physics, is referred to as the Furry representation,
yields, for the probability of $Z^0$-boson decay, an
expression in which electromagnetic interactions are
taken into account exactly, beyond standard
perturbation theory in the electromagnetic coupling
constant $\alpha=e^2/4\pi\simeq 1/137$. The explicit
form of the wave functions for charged particles in an
external electromagnetic field depends on the field
configuration. In the present study, we restrict
ourselves to the crossed-field configuration, in which
case the strength tensor $F_{\mu\nu}=\partial_\mu
A_\nu-\partial_\nu A_\mu$ has the form
\begin{equation}
\label{crossed-tensor} F_{\mu\nu}=C(k_{\mu} a_{\nu}-k_{\nu}
a_{\mu}).
\end{equation}
It is assumed in this case that the
external-electromagnetic-field potential $A_{\mu}(x)$
is chosen in the gauge
\begin{equation}
\label{gauge}
A_\mu (x)=a_\mu C(k_\nu x^\nu),
\end{equation}
Here, the unit constant $4$-vector $a_\mu$ determines
the spatial field configuration and satisfies the
conditions
\begin{equation}
\label{a-cond}
a_\mu a^\mu =-1, \qquad F_{\mu\nu}=(a_\mu F_{\nu\lambda}-a_\nu
F_{\mu\lambda})a^\lambda.
\end{equation}
A crossed field is a particular case of the field of a
plane electromagnetic wave having an isotropic wave
$4$-vector $k_\mu \ (k^2=0)$. The invariant parameter
$C$ characterizes the crossed-field strength and has
the dimensions of energy. Recall that, in an arbitrary
reference frame, a crossed field is a superposition of
constant electric and magnetic fields whose strength
vectors are orthogonal and are equal in absolute
value: $E=H=k_0 C$. Moreover, the two relativistic
invariants of the electromagnetic field are zero:
\begin{equation}
\label{crossed}F_{\mu\nu} F^{\mu\nu}=F_{\mu\nu}\tilde
F^{\mu\nu}=0.
\end{equation}
In the crossed-field model, the choice of wave
$4$-vector $k_\mu$ is quite arbitrarily. It is only
necessary that this vector be isotropic, have
dimensions of energy, and satisfy the conditions
\begin{equation}\label{k-cond}
k_\mu a^\mu =0, \qquad F^{\mu\lambda}F_{\lambda\nu}=C^2k^\mu
k_\nu.
\end{equation}
In particular, the wave $4$-vector $k^\mu$ can be
expressed in terms of the
electromagnetic-field-strength tensor and the constant
electromagnetic-potential $4$-vector $a_\lambda$ as
\begin{equation}
\label{k1}k^{\mu}=-\frac{1}{C} \cdot F^{\mu\lambda}a_{\lambda} .
\end{equation}
In dealing with the problem of $Z^0$-boson decay, it
is advisable to express the wave vector $k^\mu$ in
terms of the $Z^0$-boson 4-momentum $q^\nu$ and the
electromagnetic-field-strength tensor $F^{\mu\lambda}$
(see next section) as
\begin{equation}
\label{wavevector}k^{\mu}=\frac{e^2 \Delta}{2 M_Z^6 \varkappa^2}
\cdot F^{\mu\lambda}F_{\lambda\nu}q^\nu .
\end{equation}
In an arbitrary constant uniform electromagnetic
field, the probability of the decay $Z^0 \rightarrow
\ell^+\ell^-$ is a function of three invariant
dimensionless parameters, $P(Z^0 \rightarrow
\ell^+\ell^-)=P(\varkappa, a, b)$. These parameters
are determined by the strength tensor of an external
macroscopic field as
\begin{eqnarray}
\label{invariantk}\varkappa=\frac{e}{M_Z^3}
\sqrt{-(F_{\mu\nu}q^\nu)^2} , \\ \label{invarianta}
a=-\frac{e^2}{4 M_Z^4}F_{\mu\nu}\tilde F^{\mu\nu} ,\\
\label{invariantb} b=\frac{e^2}{4 M_Z^4} F_{\mu\nu} F^{\mu\nu} .
\end{eqnarray}
In the crossed-field model, the parameter $\varkappa$
has the simplest form
\begin{equation}
\label{kappa}
 \varkappa=\frac{e C}{M_Z^3}(q_\mu
k^\mu),
\end{equation}
the square of the wave vector (\ref{wavevector})
having the form
\begin{equation} \label{k2}
k^2=\frac{\Delta^2}{4M_Z^2}\left({a^2\over
\varkappa^4} -{2 b\over \varkappa^2}\right).
\end{equation}
In a crossed field, the two parameters
(\ref{invarianta}) and (\ref{invariantb}) of the
external-field strength vanish by virtue of the
equality in (\ref{crossed}), $a=b=0$; therefore, the
right-hand side of Eq.(\ref{k2}) also vanishes. For a
constant uniform electromagnetic field of general
form, the condition that the wave vector in
(\ref{wavevector}) is isotropic is not satisfied:
$k^2\ne 0$. Even in this case, however, we can use the
semiclassical crossed-field model, which ensures a
satisfactory description of the decay probability in
an electromagnetic field of general form, provided
that $a\ll\varkappa^2$ and $b\ll\varkappa^2 $. The
closer to zero the right-hand side of Eq.(\ref{k2})
and the higher the precision to which the condition of
isotropicity is satisfied, the more precise the
results that are obtained in this approximation. The
above constraints are obviously satisfied in the
region of relatively weak electromagnetic fields for
$a\ll 1$ and $b\ll 1$. Thus, the decay probability
$P(Z^0 \rightarrow \ell^+\ell^-)$ in a crossed field
will correspond to the first leading term $P_{0}$ in
the semiclassical expansion of the total decay
probability $P(\varkappa, a, b)$ in a power series in
the vectors of the external-field strengths at small
values of the parameters $a$ and $b$; that is,
\begin{equation}
\label{approximation} P(\varkappa, a, b)=P_{0}(\varkappa, 0, 0)
+ a \frac{\partial P}{\partial a} (\varkappa, 0, 0)+ b
\frac{\partial P}{\partial b}(\varkappa, 0, 0) + \ldots
\end{equation}
This relation between the probabilities of quantum
processes in a crossed field and a constant
electromagnetic field of general form was obtained for
the first time within quantum electrodynamics
\cite{Ritus-79}.

The wave functions for charged fermions in the field
of a plane electromagnetic wave were obtained as far
back as 1937 by D.M.Volkov (see,for example,
\cite{Landau-Lifshits-4}). In the case of a crossed
field, these exact solutions to the Dirac equations
for leptons $\ell^{\pm}$ in an external field have the
form
\begin{eqnarray}
\label{lepton-ell-} \Psi_{\ell^-}(x,p)=\exp\biggl[-ipx +
\frac{ieC(pa)}{2(pk)}(kx)^2
-\frac{ie^2C^2}{6(pk)}(kx)^3\biggr]\times \nonumber\\
\times\biggl\{1-\frac{e(kx)}{4(pk)}(F_{\mu\lambda}\gamma^\mu
\gamma^\lambda) \biggr\} \frac{u(p)} {\sqrt{2p_0 V}};
\end{eqnarray}

\begin{eqnarray}
\label{lepton-ell+} \Psi_{\ell^+}^c(x,p')=\exp\biggl[ip'x +
\frac{ieC(p'a)}{2(p'k)}(kx)^2
-\frac{ie^2C^2}{6(p'k)}(kx)^3\biggr]\times \nonumber\\
\times\biggl\{1+\frac{e(kx)}{4(p'k)}(F_{\mu\lambda}\gamma^\mu
\gamma^\lambda) \biggr\} \frac{u^c(p')} {\sqrt{2p'_0 V}};
\end{eqnarray}

The spin part of these wave functions normalized to a
three-dimensional spatial volume $V$ is determined by
the constant Dirac bispinors $u(p)$ and $u^c(p')$. The
antilepton wave function $\Psi_{\ell^+}(x, p')$
differs from expression (\ref{lepton-ell-}) only by
the reversal of the sign of the electric charge $e$:
$e\rightarrow -e$. In the expression for the matrix
element of the decay in (\ref{SZ-matrix}), the
antilepton wave function appears in the
charge-conjugate form (\ref{lepton-ell+}), which can
be derived on the basis of the Dirac-conjugate
bispinor $\bar \Psi_{\ell^+}(x, p')$ with the aid of
the charge-conjugation matrix
$U_C=-i\gamma^0\gamma^2$; that is,
\begin{equation}
\label{C-conjugate}
\Psi^c_{\ell^+}(x, p')=U_C\bar
\Psi^T_{\ell^+}(x, p')
\end{equation}
As for the $Z^0$-boson, its wave function does not
change in an electromagnetic field because it is
electrically neutral:
\begin{equation}
\label{Z-function} Z_{\mu}(x,q)=\exp\biggl(-iqx\biggr)
\frac{v_{\mu}(q)}{\sqrt{2q_0 V}}.
\end{equation}
The spin states of the $Z^0$-boson are characterized
by the complex polarization $4$-vector $v_{\mu}(q)$
satisfying standard conditions for massive vector
fields; that is,
\begin{eqnarray}
\label{Z-polarization} v_{\mu}(q) q^\mu=0, \quad v^{*}_\mu(q)
v^{\mu}(q)=-1, \\ \nonumber\sum_{\sigma=1}^3 v_{\mu}(q,\sigma)
v^*_{\nu}(q,\sigma)= - g_{\mu\nu} + q_\mu q_{\nu}/M_Z^2.
\end{eqnarray}
We substitute the particle wave functions
(\ref{lepton-ell-}), (\ref{lepton-ell+}), and
(\ref{Z-function}) into expression (\ref{SZ-matrix})
for the $S$-matrix element and perform integration of
$\mid S_{fi}\mid^2$ over the phase space of final
leptons. Upon summation over the spin states of the
lepton-antilepton pair and averaging over
polarizations of the $Z^0$-boson, we arrive at
\begin{eqnarray}
\label{Z-decay} P(Z^0\rightarrow\ell^+\ell^-\mid\varkappa)
=\frac{(g_V^2+g_A^2)M_Z^2}{12\pi^2 q_0} \int\limits_0^1 du
\biggl\{ \biggl[1-(1+3\lambda)\frac{m_\ell^2}{M_Z^2}\biggr]
\Phi_1(z)-\nonumber\\
-\frac{2\varkappa^{2/3}}{\left[u(1-u)\right]^{1/3}}
\biggl[1-2u+2u^2+(1+\lambda) \frac{m_\ell^2}{M_Z^2}\biggr]
\Phi'(z) \biggr\}.
\end{eqnarray}
The probability of $Z^0$-boson decay is expressed in
terms of the Airy functions $\Phi'(z)$ and $\Phi_1(z)$
(for necessary details concerning these special
mathematical functions, see the Appendix) depending on
the argument
\begin{equation}
\label{z} z=\frac{m_\ell^2 - M_Z^2 u(1-u)}{M_Z^2 [\varkappa u
(1-u)]^{2/3}}.
\end{equation}
The couplings of the $Z^0$-boson to the charged
leptons of all three generations, $e^{\pm},
 \mu^{\pm},$ and $ \tau^{\pm}$, can be expressed in
terms of the Weinberg angle $\theta_{\rm W}$ and the
electroweak coupling constant $g=e/\sin\theta_{\rm W}$
as
\begin{equation}
\label{gVgA} g_V=-\frac{g(1-4\sin^2\theta_{\rm
W})}{4\cos\theta_{\rm W}}, \qquad g_A=-\frac{g}{4\cos\theta_{\rm
W}} .
\end{equation}
The dimensionless parameter $\lambda$ in expression
(\ref{Z-decay}) is given by
\begin{equation}
\label{lambda}
\lambda=\frac{g_A^2-g_V^2}{g_A^2+g_V^2}=\frac{1-(1-4\sin^2\theta_{\rm
W})^2}{1+(1-4\sin^2\theta_{\rm W})^2},
\end{equation}
at $\sin^2\theta_{\rm W}=0,23$, it is very close to
unity ($\lambda \simeq0,987$).

\section{KINEMATICS OF $Z^0$-BOSON DECAY\\ IN A CROSSED FIELD
}

\begin{figure}[t]
\setlength{\unitlength}{1cm}
\begin{center}
\epsfxsize=12.cm \epsffile{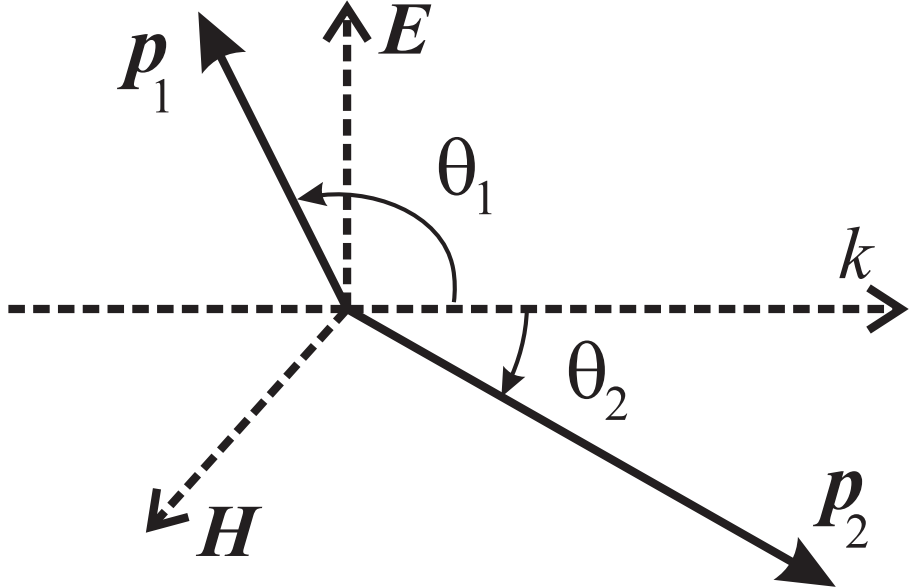}
\begin{minipage}[t]{15 cm}
\caption[]{Kinematics of the decay
$Z^0\rightarrow\ell^+\ell^-$ in a crossed field in the
$Z^0$-boson rest frame.}\label{Vectors}
\end{minipage}
\end{center}
\end{figure}

The fact that the law of energy-momentum conservation
for particles involved in the decay
$Z^0\rightarrow\ell^+\ell^-$ does not have a
conventional form,
\begin{equation}\label{law}
q_\mu + k_\mu=p_\mu+p'_\mu.
\end{equation}
is a distinctive feature of processes occurring in an
external electromagnetic field. Along with the
$Z^0$-boson and final-lepton energy-momentum
$4$-vectors $q_\mu$, $p_\mu(\ell^-)$, and
$p'_\mu(\ell^+)$, expression (\ref{law}) also involves
the wave vector (\ref{wavevector}) determining the
energy contribution of the external field. It should
be noted that, in this case, we are dealing with
asymptotic momenta that charged particles possess at
rather large distances, where the effect of the
external field is negligible, rather than with their
dynamical $4$-momenta in a crossed field. This is the
meaning that the parameters $p_\mu$ and $p'_\mu$
appearing in the Volkov solutions (\ref{lepton-ell-})
and (\ref{lepton-ell+}) to the Dirac equation in a
crossed field have. It is indeed extremely difficult
to observe the kinematics of the decay
$Z^0\rightarrow\ell^+\ell^-$ in the field itself
because the charged-lepton $4$-momenta are not
integrals of the motion. As soon as the leptons escape
from the region of the external field, their energies
and momenta do not change any longer, so that one can
measure them experimentally. In this case, the
$4$-momenta of all particles satisfy the ordinary
kinematical conditions
\begin{equation}\label{mass-shell}
p^2=p'^2=m_\ell^2, \quad q^2=M_Z^2, \quad k^2=0,
\end{equation}
and this makes it possible to represent the isotropic
vector $k^\mu$ in the form
\begin{equation}\label{k3}
k^{\mu}=\frac{\Delta}{2(q_\alpha
F^{\alpha\beta}F_{\beta\sigma}q^\sigma)}\cdot
F^{\mu\lambda}F_{\lambda\nu}q^\nu,
\end{equation}
where the parameter $\Delta$ characterizes the energy
imbalance of the reaction $Z^0\rightarrow\ell^+\ell^-$
in the external field,
\begin{equation}\label{Delta}
\Delta=2(qk)=(p+p')^2-q^2=2 m_\ell^2+2(pp')-M_Z^2
\end{equation}
Expression (\ref{law}) arises in a natural way as the
argument of the four-dimensional Dirac delta function
in calculating the $S$-matrix element
(\ref{SZ-matrix}). In performing integration over the
phase space of final leptons, there also occurs
summation of the external-field contributions
characterized by various values of the parameter
$\Delta$. This is the circumstance that distinguishes
the kinematics of $Z^0$-boson decay in an external
field from a similar process involving a real photon,
$\gamma + Z^0\rightarrow\ell^+\ell^-$, where the
$4$-momentum $k_\mu$ is fixed and is independent of
the final-lepton momenta. The physical meaning of the
above formulas can most easily be understood in the
$Z^0$-boson rest frame, where the law of
energy-momentum conservation in (\ref{law}) can be
expressed in terms of the following $4$-vectors:
\begin{equation}\label{SCM}
q^\mu=(M_Z;{\bf 0}), \quad p^\mu=(\varepsilon_1;{\bf
p}_1), \quad p'^\mu=(\varepsilon_2;{\bf p}_2), \quad
k^\mu=(k^0;{\bf k}).
\end{equation}
We chose the axes of the three-dimensional Cartesian
coordinate system in such a way that the vectors of
the electric- and magnetic-field strength, ${\bf E}$
and ${\bf H}$, are directed along the $x$ and $y$
axes, respectively. The direction of the $3$-momentum
${\bf k}$ formed by the spatial components of the wave
$4$-vector $k^\mu$ (\ref{k3}) then coincides with the
positive direction of the $z$ axis, which is parallel
to the vector product ${\bf [E\times H]}$. In the
Cartesian coordinate system chosen above, the
orientation of the lepton ($\ell^-$) and antilepton
($\ell^+$) $3$-momenta ${\bf p}_1$ and ${\bf p}_2$
with respect to the $z$ axis can be described by two
azimuthal angles $\vartheta_1$ and $\vartheta_2$
($0\le\vartheta_{1,2}\le\pi$). Figure \ref{Vectors}
shows schematically the three-dimensional
configuration formed by all of the aforementioned
vectors.

If the decay $Z^0\rightarrow\ell^+\ell^-$ occurs in
the vacuum and if the external electromagnetic field
is inoperative, the lepton and antilepton have momenta
that are equal in absolute value but are oppositely
directed and which can be expressed in terms of their
masses and the $Z^0$-boson mass. The total energy
taken away by the lepton-antilepton pair is shared
between the particles in equal parts and its total
amount is equal to the rest energy of the $Z^0$-boson,
\begin{equation}\label{F=0}
p_1=p_2=\frac{M_Z}{2}\left(1-{4 m_\ell^2\over
M_Z^2}\right)^{1/2}, \quad \varepsilon_1 =\varepsilon_2=M_Z/2,
\quad \vartheta_1+\vartheta_2=\pi.
\end{equation}
In a crossed field, each of the equalities in
(\ref{F=0}) is in general violated, and the energy
$\varepsilon_1$ that is taken away in the decay by the
lepton is not equal to the antilepton energy
$\varepsilon_2$. Only fulfillment of the following
conservation laws is guaranteed:
\begin{eqnarray}\label{SCM1}
\varepsilon_1 + \varepsilon_2 - M_Z=
p_1\cos\vartheta_1+p_2\cos\vartheta_2 , \\\label{SCM2}
p_1\sin\vartheta_1-p_2\sin\vartheta_2=0,
\end{eqnarray}
The vectors ${\bf p}_1$, ${\bf p}_2$, and ${\bf k}$
lie in the same plane, but, in this case, it is not
mandatory that the lepton and the antilepton fly apart
along the same straight line,
$\vartheta_1+\vartheta_2\ne\pi$. The deficit in the
ordinary law of energy-momentum conservation is
compensated in an external field by the isotropic wave
vector $k^\mu$ (\ref{wavevector}).

Formula (\ref{SCM1}) represents the three-dimensional
form of the conservation law of the covariant
projections of the $4$-momenta of all particles on to
the direction of the wave vector $(qk)=(pk)+(p'k)$,
while relation (\ref{SCM2}) is the conservation law
for the ordinary three-dimensional projections of the
particle momenta on to the axis that is orthogonal to
the vector ${\bf k}$ and which lies in the plane
spanned by the $3$-vectors ${\bf p}_1$, ${\bf p}_2$,
and ${\bf k}$.

\begin{figure}[t]
\setlength{\unitlength}{1cm}
\begin{center}
\epsfxsize=12.cm \epsffile{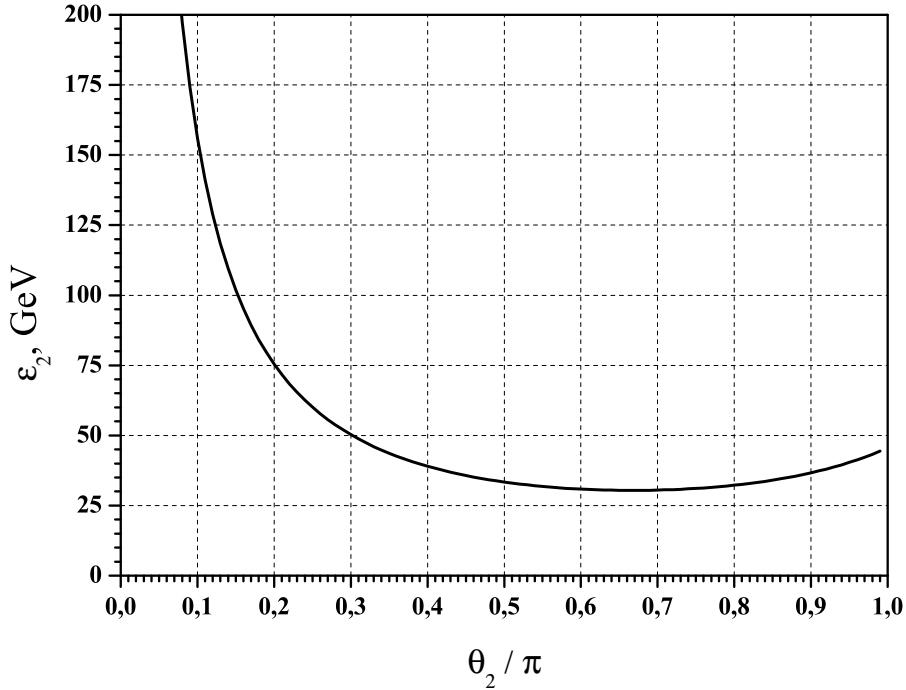}
\begin{minipage}[t]{15 cm}
\caption[]{ Antilepton energy $\varepsilon_2$ versus
the antilepton emission angle $\vartheta_2$ with
respect to the $z$ axis at a fixed value of the lepton
emission angle ($\vartheta_1=2\pi/3$).}
\label{Epsilon-2}
\end{minipage}
\end{center}
\end{figure}

We can now interpret physically the variable $u$
appearing in the integrand on the right-hand side of
(\ref{Z-decay}) and in the argument of the Airy
functions in (\ref{z}). This variable is associated
with the angles $\vartheta_1$ and $\vartheta_2$ of
divergence of the lepton-antilepton pair with respect
to the $z$ axis in the rest frame of the initial
$Z^0$-boson:
\begin{equation}\label{u}
u=\frac{(pk)}{(qk)}=\frac{\varepsilon_1-p_1\cos\vartheta_1}{M_Z}=
1-\frac{\varepsilon_2 -p_2\cos\vartheta_2}{M_Z}
\end{equation}
If the decay $Z^0\rightarrow\ell^+\ell^-$ is
unaffected by an external field, then, by virtue of
relations (\ref{F=0}), the variable $u$ can be
represented in the form
\begin{equation}\label{u0}
u={1\over 2}\left[1-\left(1-{4 m_\ell^2\over
M_Z^2}\right)^{1/2}\cos\vartheta_1\right]={1\over
2}\left[1+\left(1-{4 m_\ell^2\over
M_Z^2}\right)^{1/2}\cos\vartheta_2\right].
\end{equation}
The interval of its values admissible from the point
of view of decay kinematics in a vacuum is determined
by the inequality
\begin{equation}\label{u-limits}
u_1 \le u \le u_2,\quad \mbox{where} \quad
u_{1,2}={1\over 2}\left[1\mp\left(1-{4 m_\ell^2\over
M_Z^2}\right)^{1/2}\right]
\end{equation}
The limiting transition $\varkappa\rightarrow 0$ in
(\ref{Z-decay}) reproduces exactly well-known results
for the ordinary probability of the decay
$Z^0\rightarrow\ell^+\ell^-$ without any field. The
kinematically allowed region of values of the variable
$u$ (\ref{u-limits}) corresponds to negative values of
the argument of the Airy functions in (\ref{z});
therefore, the limiting transition to zero values of
the external-electromagnetic-field strength is
implemented by means of the formal substitution of the
Heaviside step $\theta$-function for the Airy function
$\Phi_1(z)$:
\begin{equation}\label{replace}
\Phi_1(z)\rightarrow\pi\theta(-z)
\end{equation}
In the limit $\varkappa\rightarrow 0$, the
differential probability of $Z^0$-boson decay with
respect to the variable $u$ does not depend on the
direction of divergence of leptons and has the form
\begin{equation}
\label{dZ0width-du} \frac{dP}{du}(Z^0 \rightarrow
\ell^-\ell^+)=\frac{M_Z^2}{12\pi q_0}\left[g_V^2+g_A^2+{2
m_\ell^2\over M_Z^2} (g_V^2-2 g_A^2)\right]
\end{equation}
As a matter of fact, integration with respect to the
invariant variable $u$ now reduces to multiplying
expression (\ref{dZ0width-du}) by the phase space of
this variable, $(u_2-u_1)$. As a result, we obtain the
well-known formula for the leptonic-decay width of the
$Z^0$-boson in a vacuum,
\begin{equation}\label{Zll-freewidth}
\Gamma(Z^0\rightarrow \ell^+\ell^-)=\frac{G_{\rm F}
M_Z^3}{12\pi\sqrt{2}}\sqrt{1-4\delta_\ell^2}\biggl(1-c_\ell-\delta_\ell^2(1+2
c_\ell)\biggr),
\end{equation}
where we have introduced the notation
\begin{equation}\label{ell-const}
\delta_\ell={m_\ell\over M_Z}, \qquad c_\ell=4 \sin^2
\theta_{\rm W}-8 \sin^4\theta_{\rm W}\approx 0,497
\end{equation}
In a crossed electromagnetic field, the kinematics of
$Z^0$-boson decay changes, first of all, owing to the
increase in the phase space of possible states of
final leptons $\ell^\pm$. In the three-dimensional
space of the momenta $(p_x,p_y,p_z)$ of one of the
leptons --- for example, $\ell^-$ --- the
kinematically allowed region of the decay
$Z^0\rightarrow\ell^+\ell^-$ in a vacuum [see
(\ref{F=0})] can be represented as a three-dimensional
sphere, each point of this sphere corresponding to one
of the final states of this lepton. In an external
electromagnetic field, all possible final states of
the lepton $\ell^-$ already fill some volume in
momentum space. At a fixed value of the azimuthal
angle $\vartheta_1$, the momentum of the lepton
$\ell^-$ can take any values from the interval
\begin{eqnarray}\label{pmax}
0\le p_1 < p_{\rm max}(\vartheta_1), \quad
\mbox{where} \nonumber \\ \ p_{\rm
max}(\vartheta)=\frac{M_Z}{\sin^2\vartheta}\left(\cos\vartheta
+\sqrt{1-\delta_\ell^2\sin^2\vartheta}\right).
\end{eqnarray}
The energy $\varepsilon_1$ of this particle will be
bounded by the inequality
\begin{eqnarray}\label{Emax}
m_\ell\le \varepsilon_1 <\varepsilon_{\rm
max}(\vartheta_1),\quad \mbox{where} \nonumber \\
\varepsilon_{\rm
max}(\vartheta)=\frac{M_Z}{\sin^2\vartheta}\left(1+\cos\vartheta
\sqrt{1-\delta_\ell^2\sin^2\vartheta}\right).
\end{eqnarray}
The variables $\varepsilon_1$ and $\vartheta_1$
determine unambiguously decay kinematics in a crossed
field. The energy $\varepsilon_2$ and the azimuthal
angle $\vartheta_2$ of the antilepton $\ell^+$ can be
expressed in terms of these variables with the aid of
the kinematical relations (\ref{SCM1}) and
(\ref{SCM2}) and the standard relativistic relations
\begin{equation}\label{energies12}
\varepsilon_1=\sqrt{p_1^2+m_\ell^2}, \quad
\varepsilon_2=\sqrt{p_2^2+m_\ell^2}.
\end{equation}
After some simple algebra, we obtain
\begin{eqnarray}
\label{E2exact} \varepsilon_2 = M_Z-\varepsilon_1+\frac{M_Z
(\varepsilon_1-M_Z/2)}{M_Z-\varepsilon_1+p_1\cos\vartheta_1}\\
\label{angle2exact}\vartheta_2=\arccos\left(\frac{\varepsilon_1+
\varepsilon_2-M_Z-p_1\cos\vartheta_1}{\sqrt{\varepsilon_2^2-m_\ell^2}}
\right)
\end{eqnarray}
An analysis of the resulting formulas shows that, as a
rule, the leptons $\ell^\pm$ diverge at arbitrary
angles $\vartheta_1$ and $\vartheta_2$ with respect to
the $z$ axis. This is true for almost all of the
azimuthal angles, with the exception of the case where
the lepton and antilepton move in the direction
parallel to the $z$ axis. In the case of
$\sin\vartheta_1\ne 0$ and $\sin\vartheta_2\ne 0$, the
lepton and antilepton energies $\varepsilon_1$ and
$\varepsilon_2$, respectively, can be expressed in
terms of the emission angles $\vartheta_1$ and
$\vartheta_2$ as
\begin{eqnarray}
\label{E1} \varepsilon_1\approx p_1\simeq
\frac{M_Z\sin\vartheta_2}{\sin\vartheta_1+\sin\vartheta_2-
\sin(\vartheta_1+\vartheta_2)}\\ \label{E2}\varepsilon_2\approx
p_2\simeq\frac{M_Z\sin\vartheta_1}{\sin\vartheta_1+\sin
\vartheta_2-\sin(\vartheta_1+\vartheta_2)}
\end{eqnarray}
\begin{figure}[t]
\setlength{\unitlength}{1cm}
\begin{center}
\epsfxsize=12.cm \epsffile{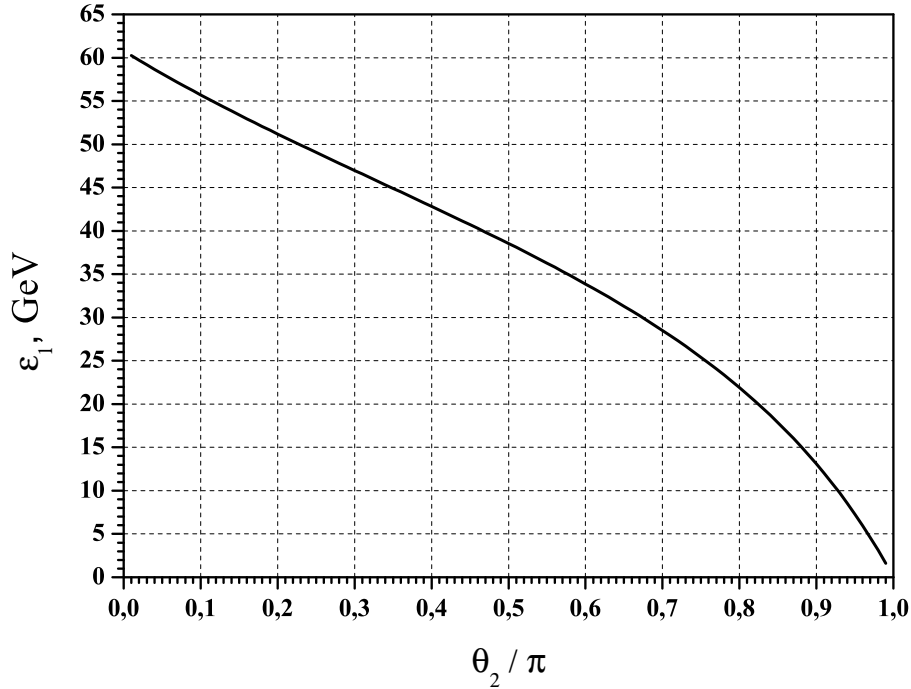}
\begin{minipage}[t]{15 cm}
\caption[]{Lepton energy $\varepsilon_1$ as a function
of the antilepton emission angle $\vartheta_2$ with
respect to the $z$ axis at $\vartheta_1=2\pi/3$.}
\label{Epsilon-1}
\end{minipage}
\end{center}
\end{figure}
These relations were obtained in the relativistic
approximation, where $\varepsilon_1 \gg m_\ell$,
$\varepsilon_2 \gg m_\ell$ and where the parameter
$\delta_\ell$ (\ref{ell-const}) can be disregarded. We
can now draw qualitative conclusions on a typical
behavior of the energies of the lepton-antilepton pair
almost over the entire azimuthal-angle range
$0<\vartheta_{1,2}<\pi$. It was indicated above that,
at a fixed value of the emission angle
($\vartheta_1={\rm const}$) of the lepton $\ell^-$,
its energy can vary from the rest energy to the
maximum value $\varepsilon_{\rm max}(\vartheta_1)$
(\ref{Emax}). Concurrently, the emission angle
$\vartheta_2$ of the antilepton $\ell^+$ can take an
arbitrary value between zero and $\pi$. As for its
energy $\varepsilon_2$, one can see that, as the
energy $\varepsilon_1$ increases, $\varepsilon_2$
first decreases from the value
\begin{equation}\label{E0}
\varepsilon_{0}=\frac{M_Z}{2}
\frac{(1-2\delta_\ell+2\delta_\ell^2)}{(1-\delta_\ell)}
\approx\frac{M_Z}{2}\left(1-\delta_\ell\right)
\end{equation}
to its local minimum
\begin{equation}\label{E2min}
\varepsilon_{2}=\frac{M_Z \cos\left(\vartheta_1/
2\right)} {1+\cos\left(\vartheta_1/2\right)},
\end{equation}
whereupon it increases monotonically to infinity as
the angle $\vartheta_2$ tends to zero, while the
lepton energy $\varepsilon_1$ tends to its maximum
value $\varepsilon_{\rm max}(\vartheta_1)$. The above
relationship between the energies of the
lepton-antilepton pair and the directions of lepton
and antilepton emission with respect to the $z$ axis
is illustrated by the graphs in Figs. \ref{Epsilon-1}
and \ref{Epsilon-2}. It should be noted that the
minimum value of the antilepton energy $\varepsilon_2$
(\ref{E2min}) is reached at the emission angle of
$\vartheta_2 = \pi-\vartheta_1/2$, in which case the
energy of the other particle, $\ell^-$,is
\begin{equation}\label{E12min}
\varepsilon_{1}=\frac{M_Z/2}{1+\cos\left(\vartheta_1/2\right)}.
\end{equation}
If the lepton escapes in the direction parallel to the
$z$ axis, expressions (\ref{E1}) and (\ref{E2}) are
inapplicable, so that one must use the exact
kinematical relations (\ref{SCM1}) and (\ref{SCM2}) or
their consequences (\ref{E2exact}) and
(\ref{angle2exact}). In this case, two situations are
possible --- that in which the leptons fly apart in
opposite directions and that in which the leptons move
in the same direction.

In the first case (at $\vartheta_1=0$ and
$\vartheta_2=\pi$ or at $\vartheta_1=\pi$ and
$\vartheta_2=0$), the energy of the lepton moving
along the positive direction of the $z$ axis can take
any value from the rest energy $\varepsilon_{\rm
min}=m_\ell$ to infinity. The energy of the second
lepton, which flies in the opposite direction, is
virtually constant --- it ranges between the initial
value $\varepsilon_0$ (\ref{E0}) and its maximum value
\begin{equation}\label{Emax-pi}
\varepsilon_{\pi}=\lim_{\vartheta\rightarrow\pi}
\varepsilon_{\rm max}(\vartheta)=\frac{M_Z}{2}
\left(1+\delta_\ell^2\right).
\end{equation}
which is very close to $\varepsilon_0$. The case where
both particles $\ell^+$ and $\ell^-$ produced in
$Z^0$-boson decay move in the same direction opposite
to the positive direction of the $z$ axis
($\vartheta_1=\vartheta_2=\pi$) is also possible. In
this case, their energies $\varepsilon_1$ and
$\varepsilon_2$ can vary from the rest energy to the
limiting value $\varepsilon_0$ (\ref{E0}). In
addition, the following relation can be obtained for
this case from (\ref{SCM1}) in the relativistic
approximation (for $\delta_\ell\rightarrow 0$):
\begin{equation}\label{E12-pi}
\varepsilon_1 + \varepsilon_2 = \frac{M_Z}{2}.
\end{equation}

Thus, one can see that, in relation to the kinematics
of the process $Z^0\rightarrow\ell^+\ell^-$ in a
vacuum [see(\ref{F=0})], the physics of $Z^0$-boson
decay in an external electromagnetic field is richer
in the number of possible final lepton states. From
the mathematical point of view, this manifests itself
in the fact that all values of the invariant variable
$u$ (\ref{u}) from the domain $0<u<1$ are now
admissible. Moreover, it is noteworthy that, in a
crossed field, the physical meaning of this variable
changes because it now characterizes two independent
kinematical quantities simultaneously - for example,
$\varepsilon_1$ and $\vartheta_1$. Disregarding the
cases in which leptons fly apart in the direction
parallel to the $z$ axis, we can express the variable
$u$ in terms of the azimuthal angles $\vartheta_1$ and
$\vartheta_2$ alone by using formulas (\ref{E1}) and
(\ref{E2}). The result is
\begin{equation}\label{uF}
u\simeq\frac{\tan({\vartheta_1/2})}
{\tan({\vartheta_1/2})+\tan({\vartheta_2/2})}.
\end{equation}

\begin{figure}[t]
\setlength{\unitlength}{1cm}
\begin{center}
\epsfxsize=12.cm \epsffile{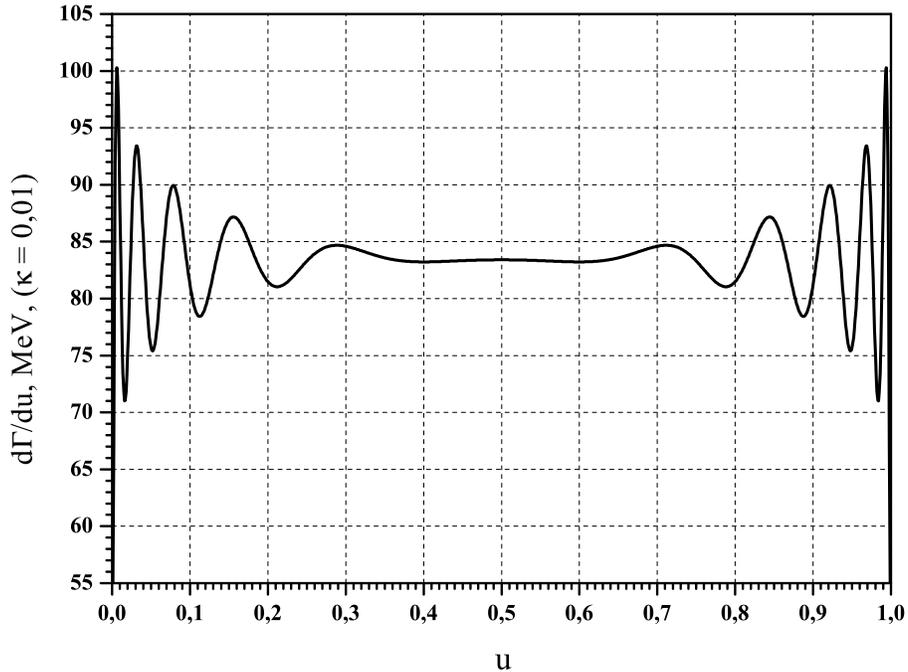}
\begin{minipage}[t]{15 cm}
\caption[]{Differential width with respect to the
decay $Z^0 \rightarrow \ell^-\ell^+$ as a function of
the invariant angular variable $u$ (\ref{u}) in a weak
crossed field (at $\varkappa=0,01$)}\label{dZdu}
\end{minipage}
\end{center}
\end{figure}
An analysis of the integrand on the right-hand side of
(\ref{Z-decay}) shows that, in an external field, not
all of the directions of divergence of the lepton and
antilepton are equiprobable. The graphs in Fig.
\ref{dZdu} that represent the dependence of the
differential decay width of the $Z^0$-boson on the
invariant variable $u$ corroborate that, even in a
relatively weak external field, there are noticeable
deviations from a uniformly distributed probability of
the leptonic decay mode in a vacuum [see
(\ref{dZ0width-du})]. In the case where leptons
produced in the process $Z^0\rightarrow\ell^+\ell^-$
fly apart at specific angles $(\vartheta_{1,2}\ne
0,\pi)$ in diametrically opposite directions
$(\vartheta_1+\vartheta_2\approx \pi)$, the invariant
variable $u$ takes values of $u\approx 0,5$. In this
region, which is kinematically allowed for $Z^0$-boson
decay in a vacuum, the deviations of the differential
decay width $\Gamma'_u(u,\varkappa)$ from the
corresponding static value of $\Gamma'_u(u,0)\approx
83$ MeV are in fact unobservable. If one of the
leptons flies away at a specific angle $\vartheta$
with respect to the $z$ axis and at an energy close to
the maximum value $\varepsilon_{\rm max}(\vartheta)$
in (\ref{Emax}) and if the second lepton moves along
the direction nearly parallel to this axis (in the
positive direction), the variable $u$ takes values in
the region around $u\approx 0$ or $u\approx 1$. In
this case, one can observe oscillations of the
differential decay width of the $Z^0$-boson that are
characterized by a growing amplitude, which increases
and decreases the static vacuum value in
(\ref{dZ0width-du}) by more than 15\%. Similar
oscillations of probabilities of quantum processes in
an external electromagnetic field are characteristic
of all particle transformations allowed in a vacuum
and were described in detail in the literature
\cite{Ritus-79,Ternov-Khalilov-Rodionov}.

\section{LEPTONIC DECAY WIDTH OF THE $Z^0$-BOSON}
\hspace*{\parindent}

The integral representation of the partial width of
the $Z^0$-boson with respect to its decay to a pair of
charged leptons in an external electromagnetic field
can be derived from expression (\ref{Z-decay}) for the
probability by going over to the rest frame
($q_0=M_Z$) and by making the change of the
integration variable $x=u(1-u)$. The result is
\begin{eqnarray}
\label{Z-width} \Gamma(Z^0\rightarrow\ell^+\ell^-\mid\varkappa)
&=&\frac{(g_V^2+g_A^2)M_Z}{6\pi^2}
\int\limits_0^{1/4}\frac{dx}{\sqrt{1-4x}} \biggl\{
\biggl[1-\delta_\ell^2(1+3\lambda)\biggr] \Phi_1(z)-\nonumber\\
&-&2\left(\frac{\varkappa^2}{x}\right)^{1/3} \biggl[
1-2x+\delta_\ell^2(1+\lambda)\biggr] \Phi'(z) \biggr\},
\end{eqnarray}
where the argument of the Airy functions is now
determined by the expression
\begin{equation}
\label{zx} z=\frac{\delta_\ell^2 - x}{(\varkappa x)^{2/3}}.
\end{equation}
Let us now consider the asymptotic estimates obtained
from this formula at various values of the
external-field-strength parameter $\varkappa$
(\ref{kappa}).

In weak electromagnetic fields such that
$\varkappa\ll\delta_\ell^2\ll 1$, the partial decay
width of the $Z^0$-boson can be written as the sum of
two terms,
\begin{equation}\label{sum}
\Gamma(Z^0\rightarrow\ell^+\ell^-
\mid\varkappa)=\Gamma(Z^0\rightarrow
\ell^+\ell^-)+\Delta\Gamma(\varkappa) .
\end{equation}
The first term in this expression is the $Z^0$-boson
width with respect to the leptonic decay
$Z^0\rightarrow \ell^+\ell^-$ in a vacuum [see
expression (\ref{Zll-freewidth}) above]. The
experimental value of this quantity is known to a
precision of 0.1\%: $\Gamma(Z^0\rightarrow
\ell^+\ell^-)=83,984\pm 0,086\ \mbox{MeV}$
\cite{PDG-2006}. The second term characterizes the
effect of the external electromagnetic field. In the
leading order of the expansion in the small parameters
$\varkappa$ and $\delta_\ell$, this term can be
represented in the form
\begin{eqnarray}\label{correction}
\Delta\Gamma(\varkappa)=-\frac{G_{\rm F}
M_Z^3}{4\pi\sqrt{6}}\Biggl[\ \delta_\ell^2 (1-6 c_\ell)
\varkappa \cos\left({1\over 3\varkappa}\right)-\nonumber\\
-(1-c_\ell) \varkappa^2 \sin\left({1\over
3\varkappa}\right)+\frac{8\varkappa^2}{\sqrt{3}}(1- c_\ell)\
\Biggr]
\end{eqnarray}
Similar calculations of the probability of $Z^0$-boson
decay in a magnetic field were performed in
\cite{Borisov-Ivuz-87}, where the contribution of the
virtual electron-positron loop to the amplitude for
elastic $Z^0$-boson scattering was considered, its
imaginary part being directly related to the
$Z^0\rightarrow\ell^+\ell^-$ decay width being
studied. It is worth noting, however, that the results
of the present calculations [expressions
(\ref{Z-width}) and (\ref{correction})] differ
somewhat from their counter parts in
\cite{Borisov-Ivuz-87}.

\begin{figure}[t]
\setlength{\unitlength}{1cm}
\begin{center}
\epsfxsize=15.cm \epsffile{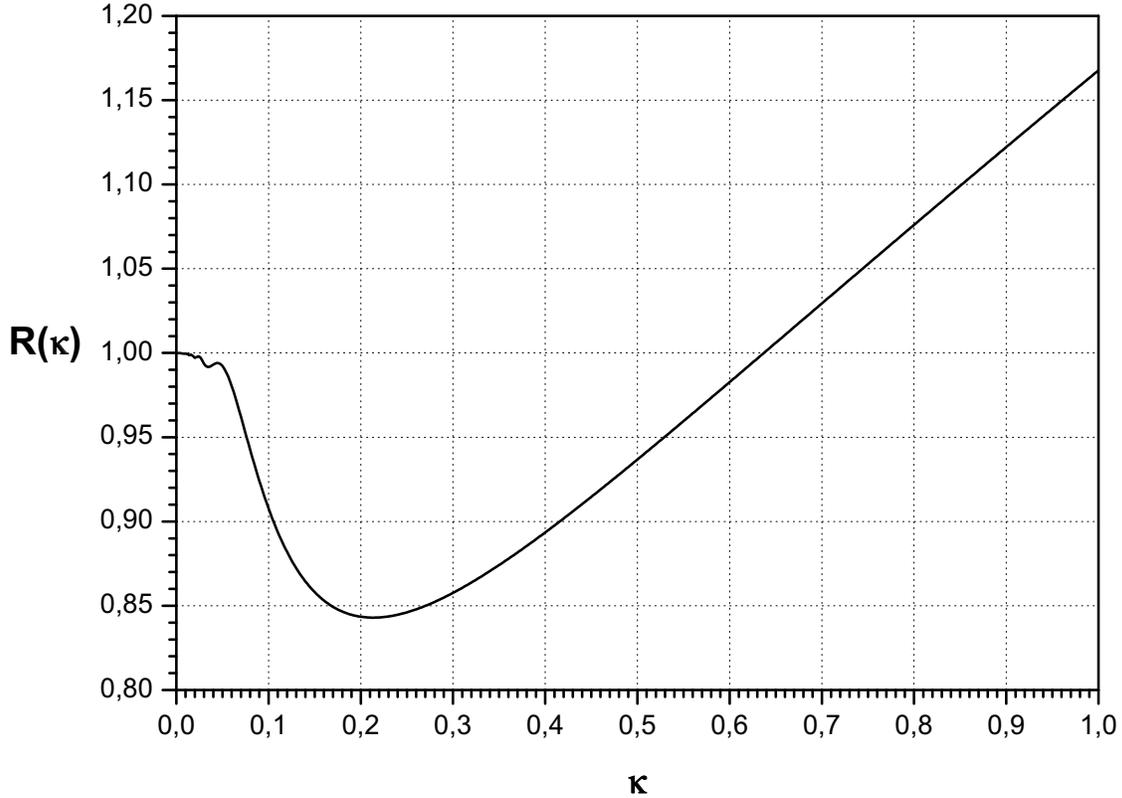}
\begin{minipage}[t]{15 cm}
\caption[]{Leptonic-decay width of the $Z^0$-boson as
a function of the external-field-strength parameter
$\varkappa$ (\ref{kappa}) in the vicinity of the
absolute-minimum point.}\label{ZR3}
\end{minipage}
\end{center}
\end{figure}

For the parameter of the
external-electromagnetic-field strength, we will now
consider the region specified by the condition
$\varkappa\gg\delta_\ell^2$. We use the fact that the
masses of the leptons of all three generations are
much smaller than the $Z^0$-boson mass and go over to
the limit $\delta_\ell\rightarrow 0$ in formulas
(\ref{Z-width}) and (\ref{zx}). We can then calculate
analytically the integral of the Airy function on the
right-hand side of (\ref{Z-width}) at arbitrary values
of the external-field-strength parameter $\varkappa$.
The final expression for the leptonic-decay width of
the $Z^0$-boson is written in terms of the Bessel
functions $J_\nu(w)$ carrying the noninteger indices
$\nu=\pm 1/6$ and $\nu=\pm 5/6$ and depending on the
argument $w=(6\varkappa)^{-1}$. We now consider a
universal function $R(\varkappa)$ that describes the
degree of influence of the external field on the
leptonic mode of $Z^0$-boson decay:
\begin{equation}\label{ratio}
R(\varkappa)=\frac{\Gamma(Z^0\rightarrow\ell^+\ell^-
\mid\varkappa)}{\Gamma(Z^0\rightarrow \ell^+\ell^-)}.
\end{equation}
\begin{figure}[t]
\setlength{\unitlength}{1cm}
\begin{center}
\epsfxsize=15.cm \epsffile{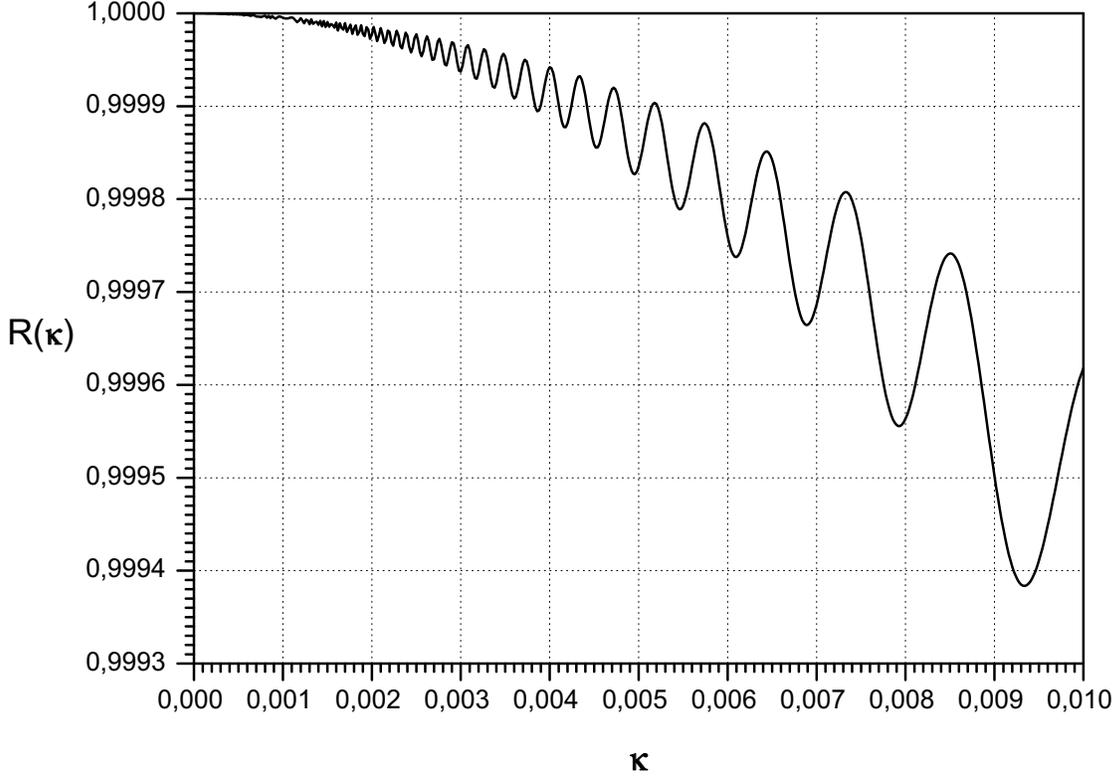}
\begin{minipage}[t]{15 cm}
\caption[]{Oscillations of the partial width of the
$Z$-boson with respect to its decay to a pair of
charged leptons in a weak electromagnetic
field.}
\label{ZR1}
\end{minipage}
\end{center}
\end{figure}
In the massless-lepton approximation ($\varkappa \gg
\delta_\ell^2$), this function, which is a relative
width of the $Z^0$-boson with respect to the decay
$Z^0\rightarrow \ell^+\ell^-$ in an electromagnetic
field, can be represented in the form
\begin{equation}
\label{Ratio-k}
R(\varkappa)=\frac{1}{3}\Biggl[1+R_1(\varkappa)+R_2(\varkappa)+
R_3(\varkappa)\Biggr],
\end{equation}
where the following notation has been introduced:
\begin{equation}
\label{R1} R_1(\varkappa)=\frac{\pi
(1+3\varkappa^2)}{12\varkappa}\left[J^2_{1/6}\left({1\over
6\varkappa}\right)+J^2_{-1/6}\left({1\over
6\varkappa}\right)\right],
\end{equation}
\begin{equation}
\label{R2} R_2(\varkappa)=\frac{\pi
(1-15\varkappa^2)}{12\varkappa}\left[J^2_{5/6}\left({1\over
6\varkappa}\right)+J^2_{-5/6}\left({1\over
6\varkappa}\right)\right],
\end{equation}
\begin{eqnarray}
\label{R3} R_3(\varkappa)&=&\frac{\pi
(2+15\varkappa^2)}{6}\Biggl[\ J_{1/6}\left({1\over
6\varkappa}\right)J_{-5/6}\left({1\over
6\varkappa}\right)-\nonumber\\ &-& J_{-1/6}\left({1\over
6\varkappa}\right)J_{5/6}\left({1\over 6\varkappa}\right)\
\Biggr].
\end{eqnarray}
\begin{figure}[t]
\setlength{\unitlength}{1cm}
\begin{center}
\epsfxsize=15.cm \epsffile{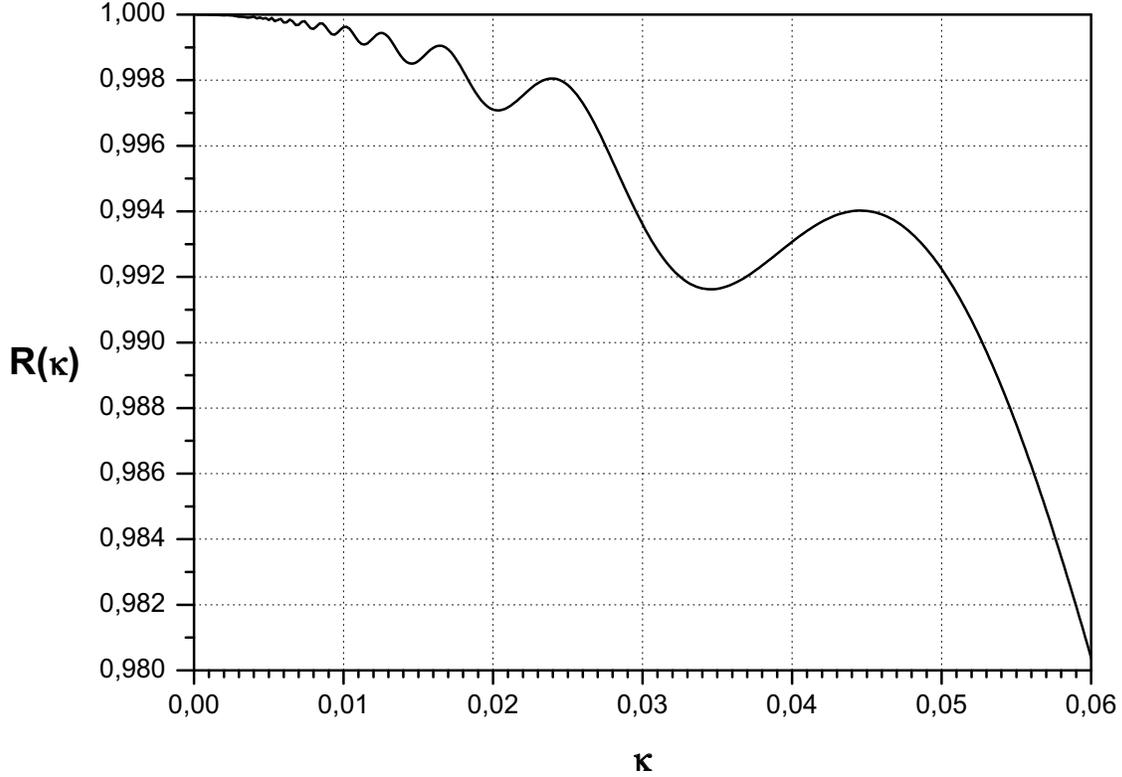}
\begin{minipage}[t]{15 cm}
\caption[]{Relative $Z^0\rightarrow\ell^+ \ell^-$
decay width $R(\varkappa)$ (\ref{ratio}) as a function
of the parameter $\varkappa$ at the boundary of the
region of oscillations.}\label{ZR2}
\end{minipage}
\end{center}
\end{figure}
The graphs representing the dependence of the relative
leptonic-decay width (\ref{ratio}) normalized to unity
at zero external field on the invariant field-strength
parameter $\varkappa$ (\ref{invariantk}) are on
display in Figs. \ref{ZR1}, \ref{ZR2}, \ref{ZR3}. In
relatively weak fields (such that $\varkappa < 0,06$),
there arise oscillations of the probability of the
decay $Z^0\rightarrow\ell^+\ell^-$, their amplitude
increasing quadratically with the parameter
$\varkappa$. In this region, one can use the
asymptotic expansion for the Bessel functions and
approximate formula (\ref{Ratio-k}) by the expression
\begin{eqnarray}
\label{RAs1}
R(\varkappa)=1-\frac{16}{3}\varkappa^2+\frac{3\varkappa^2}
{\sqrt{3}}\sin\left({1\over 3\varkappa}\right)
+\frac{65\varkappa^3} {\sqrt{3}}\cos\left({1\over
3\varkappa}\right)+\nonumber\\ + \frac{880\varkappa^4}
{3\sqrt{3}}\sin\left({1\over 3\varkappa}\right)
-\frac{160}{3}\varkappa^4 + \ldots
\end{eqnarray}
It is obvious from Figs.~\ref{ZR1} and \ref{ZR2} that
the maximum deviation of the decay width oscillating
in an external field from its vacuum value in
(\ref{Zll-freewidth}) does not exceed 2\%. A further
increase in the parameter of the
external-electromagnetic-field strength to values in
the region $\varkappa > 0,06$ leads to a complete
disappearance of oscillations and a monotonic decrease
in the relative width $R(\varkappa)$ down to the
absolute minimum $R_{\min}=0,843$ in the vicinity of
the point $\varkappa_{\rm min}=0,213$ (see Fig.
\ref{ZR3}). In the region of strong fields (such that
$\varkappa > 0,213$), the relative width of the
$Z^0$-boson with respect to its leptonic decay mode
grows monotonically with the field strength. In the
region $\varkappa\gg 1$, this growth can be
approximated by the asymptotic expression.
\begin{equation}
\label{RAs2} R(\varkappa)=\frac{15\
\Gamma^4(2/3)}{14\pi^2}(3\varkappa)^{2/3} + \frac{1}{3} +
\frac{3\ \Gamma^4(1/3)}{110\pi^2\ (3\varkappa)^{2/3}} + \ldots
\end{equation}

\section{HADRONIC MODES OF $Z^0$-BOSON DECAY
}

The results obtained in the preceding section admit a
straight forward generalization to the case of the
decay $Z^0\rightarrow \bar f f$, where, by $f$, we
denote an arbitrarily charged fermion, $q_f$ standing
for the absolute value of its electric charge. As is
well known, the charge is $q_f=2/3$ for the up quarks
($u, c, t$) and $q_f=1/3$ for the down quarks ($d, s,
b$). It is necessary to consider that, for the $Z\bar
f f$ vertex, the coupling constants $g_V$ and $g_A$
(\ref{SZ-matrix}) also depend on the absolute value of
the electric charge ($q_f$). In the
Glashow-Weinberg-Salam model of electroweak
interactions, they are determined by the relations
\begin{equation}
\label{gV-gA} \mid g_V(f)\mid=\frac{g(1-4q_f\sin^2\theta_{\rm
W})}{4\cos\theta_{\rm W}}, \qquad \mid g_A(f)\mid=
\frac{g}{4\cos\theta_{\rm W}} .
\end{equation}
Upon making the formal substitution $e\rightarrow q_f
e$ or $\varkappa\rightarrow q_f\varkappa$ in
(\ref{Z-decay}), the differential $Z^0\rightarrow \bar
f f$ decay width as a function of the invariant
variables $\varkappa$ (\ref{invariantk}) and $u$
(\ref{u}) assumes the form
\begin{eqnarray}
\label{Zff-width} \frac{d\Gamma}{du}(Z^0\rightarrow \bar f
f)=\frac{G_{\rm F} M_Z^3} {12\pi^2\sqrt{2}}
\Biggl\{\left[1-c_f-\delta_f^2 (1+2 c_f)\right]\Phi_1(z_f)-\\
\nonumber -2\left[\frac{q_f^2\varkappa^2}{u(1-u)}\right]^{1/3}
\left[(1-c_f)(1-2u+2u^2)+\delta_f^2\right] \Phi'(z_f)\Biggr\}.
\end{eqnarray}
In this expression, use is made of the same notation
as in (\ref{z}) and (\ref{ell-const}):
\begin{equation}\label{d&qF}
 \delta_f={m_f\over M_Z},  \qquad c_f=4 q_f \sin^2\theta_{\rm
W}(1-2 q_f \sin^2\theta_{\rm W}),
\end{equation}
\begin{equation}\label{z_f}
z_f=\frac{\delta_f^2 - u(1-u)}{[q_f\varkappa u (1-u)]^{2/3}}.
\end{equation}
The limiting transition to an infinitely weak external
field ($\varkappa\rightarrow 0$) leads to reproducing
the well-known results for the partial widths of the
$Z^0$-boson with respect to its decay to light quarks;
that is,
\begin{equation}\label{Zup}
\Gamma(Z^0\rightarrow \bar u u)=\frac{G_{\rm F}
M_Z^3}{4\pi\sqrt{2}} \left[1-\frac{8}{3} \sin^2\theta_{\rm
W}+\frac{32}{9} \sin^4\theta_{\rm W}\right].
\end{equation}

\begin{equation}\label{Zdown}
\Gamma(Z^0\rightarrow \bar d d)=\frac{G_{\rm F}
M_Z^3}{4\pi\sqrt{2}} \left[1-\frac{4}{3} \sin^2\theta_{\rm
W}+\frac{8}{9} \sin^4\theta_{\rm W}\right].
\end{equation}
These relations are obtained from
(\ref{Zll-freewidth}) upon the formal substitution
$c_\ell\rightarrow c_f$ and $\delta_\ell\rightarrow
\delta_f$ and the multiplication by the common factor
$N_c=3$ (number of color quark states). The inclusion
of radiative corrections leads to various
modifications of expressions (\ref{Zup}) and
(\ref{Zdown}), whose explicit form depends on the
choice of input parameters in electroweak theory and
renormalization scheme for divergent loop diagrams
\cite{UFN-96, 2-loops-tt}. For example, Sirlin, who
was the author of one of the pioneering studies in
these realms \cite{Sirlin1980}, proposed the
parametrization
\begin{equation}\label{Sirlin}
G_{\rm F}=\frac{\pi\alpha}{\sqrt{2} M_W^2 \sin^2\theta_{\rm W}}
\frac{1}{(1-\Delta r)}, \qquad \sin^2\theta_{\rm
W}=1-\frac{M_W^2}{M_Z^2},
\end{equation}
which has been widely used since then. Here, $\Delta
r$ includes all radiative corrections. For the sake of
simplicity, we restrict ourselves to the case of
massless quarks ($\delta_f\ll\varkappa$) and, without
going in to details of the calculation of radiative
corrections, try to analyze the effect of external
electromagnetic fields on the hadronic decay modes of
the $Z^0$-boson. Without allowance for the
external-field effect, the width $$
\Gamma_{had}=\sum^5_{i=1}\Gamma(Z^0\rightarrow \bar
q_i q_i)\simeq 2 \Gamma(Z^0\rightarrow \bar u u)
+3\Gamma(Z^0\rightarrow \bar d d) $$ agrees with the
experimentally measured value of
$\Gamma(Z^0\rightarrow hadrons)=1744,4\pm 2,0$ MeV
\cite{PDG-2006} to a high degree of precision. We use
this approximation in the region of relatively strong
electromagnetic fields ($\varkappa\sim 1$) and
estimate the width of the $Z^0$-boson with respect to
the hadronic decay mode as
\begin{equation}\label{Zhadrons}
\Gamma_{had}(\varkappa)=2 \Gamma(Z^0\rightarrow \bar u u)
R\left(\frac{2}{3}\varkappa\right)+3\Gamma(Z^0\rightarrow \bar d
d) R\left(\frac{1}{3}\varkappa\right),
\end{equation}
\begin{figure}[t]
\setlength{\unitlength}{1cm}
\begin{center}
\epsfxsize=15.cm \epsffile{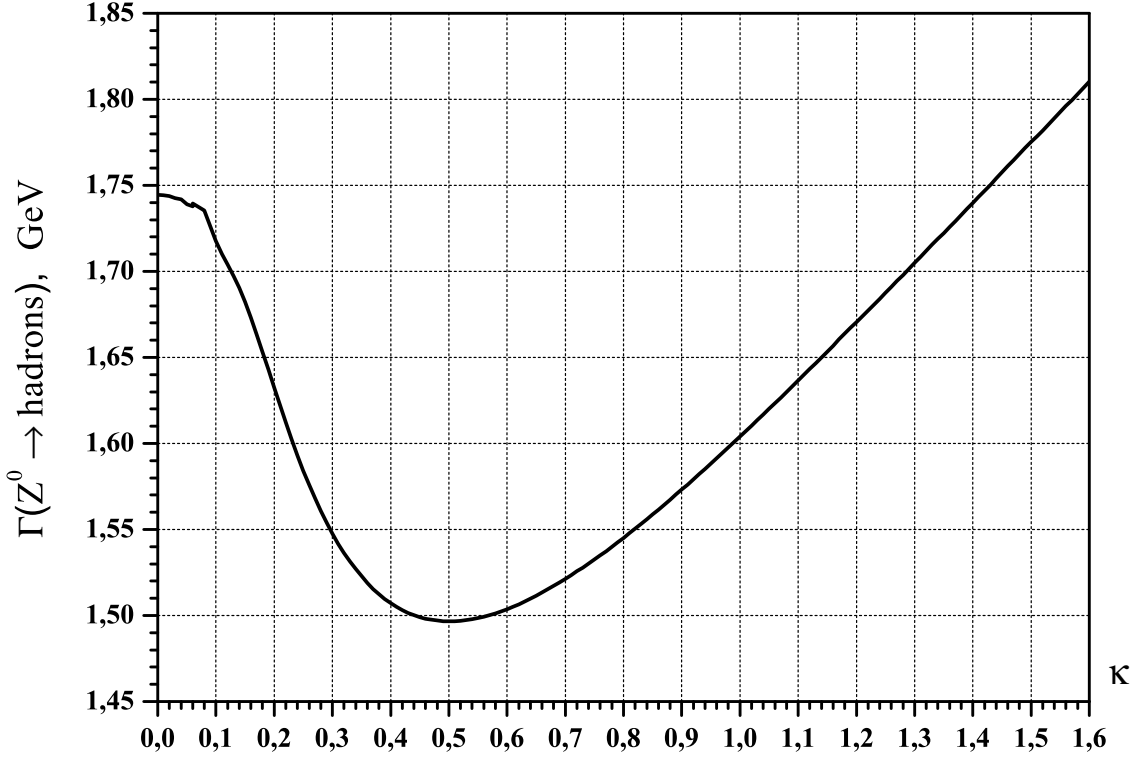}
\begin{minipage}[t]{15 cm}
\caption[]{Hadronic-decay width of the $Z^0$-boson in
an external electromagnetic field as a function of the
parameter $\varkappa$ (\ref{kappa}).}\label{Z-hadrons}
\end{minipage}
\end{center}
\end{figure}
where the function $R(\varkappa)$ is determined by
expressions (\ref{ratio}) -- (\ref{R3}). Figure
\ref{Z-hadrons} shows the results of numerical
calculations performed for the total width of the
$Z^0$-boson with respect to hadronic decay modes in an
external field on the basis of expression
(\ref{Zhadrons}) with allowance for the production of
only five quarks ($u,d,s,c,$ and $b$). In this region,
the contribution of the heavy $t$ quark is
insignificant. This contribution will be considered
individually in the next section.

One can see from Fig. \ref{Z-hadrons} that, in just
the same way as the width with respect to leptonic
decay modes (see Fig. \ref{ZR3}), the width with
respect to the hadronic modes of $Z^0$-boson decay in
an intense external field behaves nonmonotonically. As
the parameter $\varkappa$ increases, the width
$\Gamma_{had}(\varkappa)$ first decreases, reaching
the absolute minimum of $\Gamma_{had}(\varkappa_{\rm
min}) = 1497$ MeV at the point $\varkappa_{\rm
min}=0,501$, where upon it begins growing. In a weak
electromagnetic field ($\varkappa\ll 1$), the gradual
decrease in the hadronic-decay width
$\Gamma_{had}(\varkappa)$ is accompanied by quite
complicated oscillations.

\section{TOTAL DECAY WIDTH OF THE $Z^0$-BOSON AND $t$-QUARK
 CONTRIBUTION}

In a vacuum, $Z^0$-boson decay to $t$ quarks via the
process $Z^0\rightarrow \bar t t$ is forbidden by the
energy-conservation law because the mass of the heavy
$t$ quark is rather high, $m_t=174,2\pm 3,3$~GeV
\cite{PDG-2006}. The presence of an external
electromagnetic field removes this forbiddance,
opening new channels of $Z^0$-boson decay. In
principle, the production of arbitrary charged
fermions whose masses satisfy the relation
$\delta_f=m_f/M_Z > 1/2$ becomes possible. The missing
energy for these reactions comes from the external
field, and this resembles tunneling through a
potential barrier in nonrelativistic quantum
mechanics. At small values of the
electromagnetic-field strength (such that $\varkappa
\ll 1$), the $Z^0\rightarrow \bar f f$ decay width is
exponentially small:
\begin{figure}[t]
\setlength{\unitlength}{1cm}
\begin{center}
\epsfxsize=15.cm \epsffile{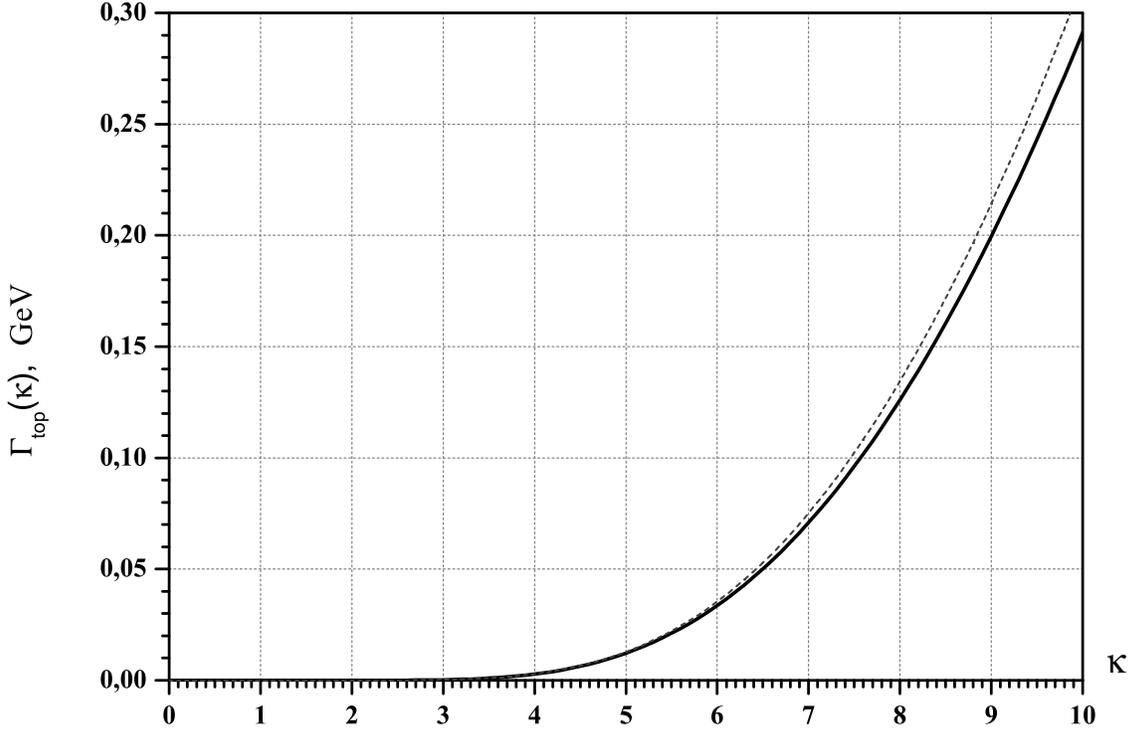}
\begin{minipage}[t]{15 cm}
\caption[]{Partial width of the $Z^0$-boson with
respect to its decay to $t$-quarks in an external
electromagnetic field (for the description of the
curves, see the main body of the text).}
\label{Z-topgraf}
\end{minipage}
\end{center}
\end{figure}
\begin{equation}
\label{Zff-massive} \Gamma(Z^0\rightarrow \bar f
f)=\frac{G_{\rm F} M_Z^3
\left(8\delta_f^2+1-6c_f\right)\delta_f^2 q_f}
{8\pi\sqrt{6}\left(4\delta_f^2-1\right)\sqrt{8\delta_f^2+1}}
\varkappa\
\exp\left[-\frac{\left(4\delta_f^2-1\right)^{3/2}}
{3q_f\varkappa}\right].
\end{equation}
This formula was derived from expression
(\ref{Zff-width}) upon integration with respect to the
invariant variable $u$ (\ref{u}) in the approximation
$z_f\gg 1$ (\ref{z_f}). Substituting the $t$-quark
charge $q_f=2/3$ and performing summation over three
color states, we obtain an asymptotic estimate for the
partial width of the $Z^0$-boson with respect to its
decay to $t$-quarks in a relatively weak
electromagnetic field; that is,
\begin{eqnarray}
\label{Z-top} \Gamma(Z^0\rightarrow \bar t t)=\frac{G_{\rm
F}m_t^2 M_Z^2 \left[8 m_t^2+\left(1 - 16\sin^2\theta_{\rm
W}+\frac{64}{3} \sin^4\theta_{\rm W}\right) M_Z^2\right]}
{4\pi\sqrt{6}\left(4 m_t^2 - M_Z^2\right)\sqrt{8 m_t^2 + M_Z^2}}
\times \nonumber\\ \varkappa\ \exp\left[-\frac{\left(4 m_t^2 -
M_Z^2 \right)^{3/2}} {2\varkappa M_Z^3}\right].
\end{eqnarray}
The results of numerical calculations show that the
above approximate formula for $\Gamma(Z^0\rightarrow
\bar t t)$ leads to an error not greater than 4\% for
all values of the external-field-strength parameter
from the region $\varkappa \le 5$. As the parameter
$\varkappa$ grows, the deviation of the results
obtained according to the asymptotic formula
(\ref{Z-top}) from the precise value of the partial
width of the $Z^0$-boson with respect to the decay
$Z^0\rightarrow \bar t t$ in an external field
increases [see Fig.\ref{Z-topgraf}, where the dotted
curve corresponds to the calculation according to
formula (\ref{Z-top}), while the solid curve
represents the results of a numerical integration of
expression (\ref{Zff-width})].
\begin{figure}[t]
\setlength{\unitlength}{1cm}
\begin{center}
\epsfxsize=15.cm \epsffile{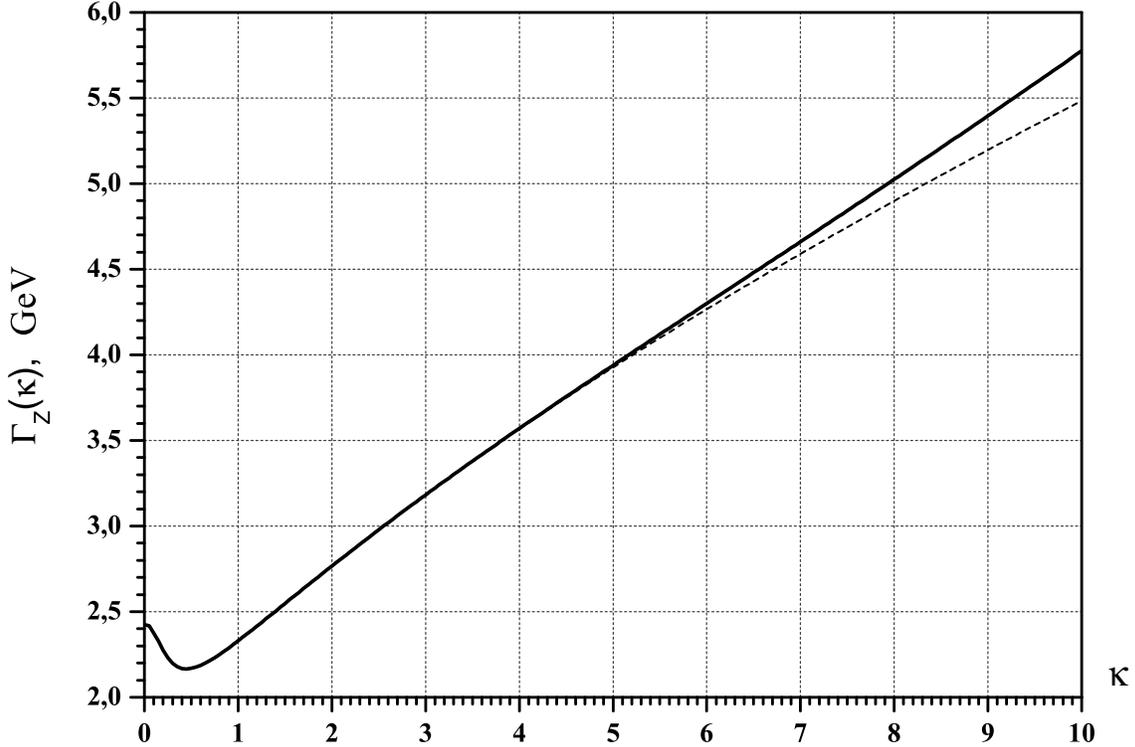}
\begin{minipage}[t]{15 cm}
\caption[]{Total decay width of the $Z^0$-boson in an
external electromagnetic field (for the description of
the curves, see the main body of the
text).}\label{Z-total-10}
\end{minipage}
\end{center}
\end{figure}

The $t$-quark contribution to the total decay width of
the $Z^0$-boson remains negligible (less than 1\%) up
to strength-parameter values of about $\varkappa =
6,4$. Figure~\ref{Z-total-10} displays the total decay
width of the $Z^0$-boson as a function of the
parameter $\varkappa$ in the region of strong fields
($\varkappa \le 10$). The solid curve corresponds to
the total decay width of the $Z^0$-boson with
allowance for the $t$-quark contribution, while the
dotted curve represents the same quantity calculated
without taking into account the $t$-quark
contribution. One can clearly see a trend toward the
growth of both the absolute value of the total decay
width of the $Z^0$-boson and the fraction of the
$t$-quark contribution in this decay width. As the
external-field strength increases, the process
$Z^0\rightarrow \bar t t$, which is forbidden in a
vacuum, becomes dominant in superstrong
electromagnetic fields. By way of example, it can be
indicated that, in the $Z^0$-boson decay width at
$\varkappa = 100$, the fraction associated with
$t$-quarks is as large as 50\%. This circumstance
clearly demonstrates how an external electromagnetic
field can change drastically the standard physics of
quantum processes in a vacuum and serve as some kind
of a catalyst for a number of new and nontrivial
phenomena.

\newpage
\section{CONCLUSIONS}

The effect of a strong electromagnetic field on the
probability of $Z^0$-boson decay has been
investigated. The present calculations have been
performed within the crossed-field model, which makes
it possible to trace characteristic variations in the
modes of $Z^0$-boson decay to known leptons and
quarks. We have found that, in the region of
relatively strong fields ($\varkappa \sim 1$), all
partial decay widths $\Gamma(Z^0\rightarrow \bar f f)$
decrease in the same manner by about $12$ to $15\%$.
This decrease can be described by a universal function
$R(\varkappa)$ [see (\ref{ratio})--(\ref{R3})]
depending only on the energy-momentum of the
$Z^0$-boson and on the external-electromagnetic-field
strength. This circumstance becomes quite obvious if
we consider that all known fermions observed in
$Z^0$-boson decays have masses at least an order of
magnitude smaller than the $Z^0$-boson mass.
Therefore, it comes as no surprise that, in the region
of strong electromagnetic fields, the massless-fermion
approximation works well, also making it possible to
calculate the dependence of the total decay width
$\Gamma_Z(\varkappa)$ of the $Z^0$-boson on the
external-field-strength parameter $\varkappa$
(\ref{invariantk}).

\begin{figure}[t]
\setlength{\unitlength}{1cm}
\begin{center}
\epsfxsize=15.cm \epsffile{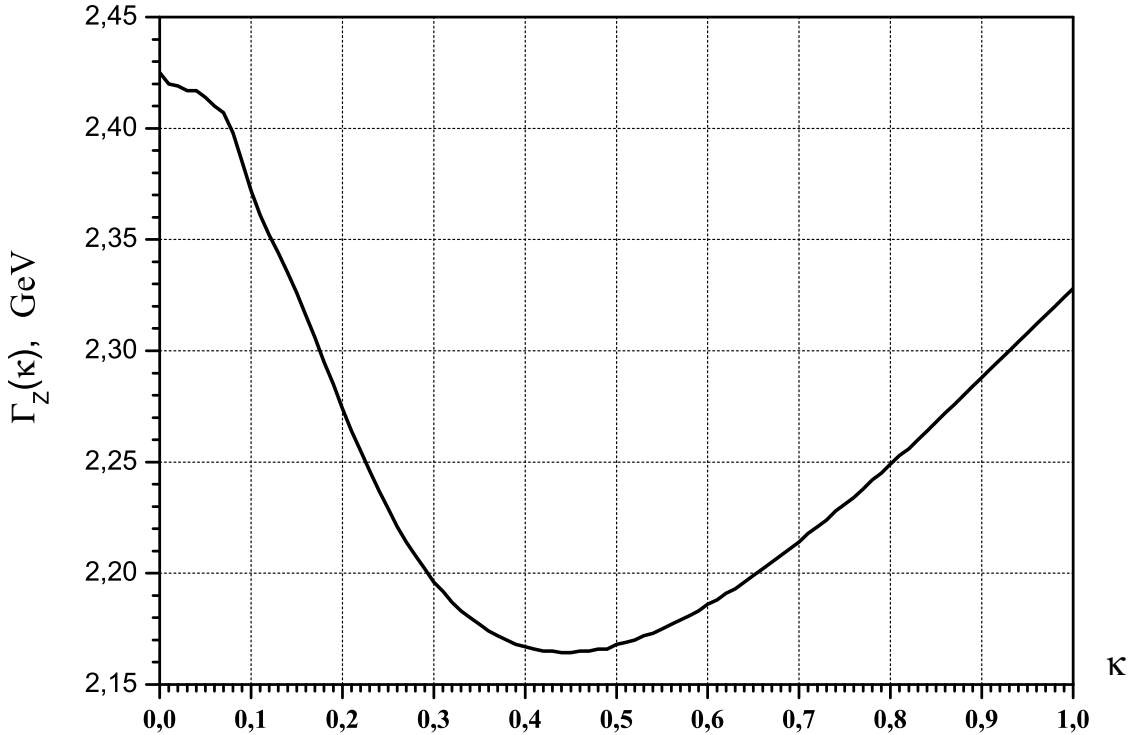}
\begin{minipage}[t]{15 cm}
\caption[]{Minimal value of the total decay width of
the $Z^0$-boson in an external field.}\label{Z-total}
\end{minipage}
\end{center}
\end{figure}

The results of the present calculations have revealed
that, in just the same way as in the case of the
$W$-boson \cite{Kurilin-2004}, the total decay width
of the $Z^0$-boson in a strong electromagnetic field
depends nonmonotonically on the strength parameter
$\varkappa$, this dependence featuring the point of a
local minimum, $\Gamma_Z(\varkappa_{\rm min}) =
2,164$~GeV, at $\varkappa_{\rm min}=0,445$. The graph
in Fig.~\ref{Z-total}, which represents the behavior
of $\Gamma_Z(\varkappa)$ in the region $\varkappa \le
1$, illustrates this observation.

For the sake of comparison, it can be indicated that
the analogous $W$-boson-decay process
$W^-\rightarrow\ell \bar\nu_\ell$ in an external field
is also characterized by a nonmonotonic dependence of
the decay width on the parameter $\varkappa$ with the
point of a local minimum at $\varkappa_{\rm W}=0,6112$
\cite{Kurilin-2004}. Upon rescaling this value to the
$Z^0$-boson energy scale, it corresponds to
$\varkappa=\varkappa_{\rm W} \left(M_W /M_Z\right)^3 =
0.419$. Thus, the probabilities for the decays of
gauge bosons in an external field, $Z^0\rightarrow
\bar f f$ and $W^-\rightarrow\ell \bar\nu_\ell$, take
their minimum values at nearly the same field
strength. This agreement is not accidental since the
two reactions in question proceed under similar
kinematical conditions and since the masses of the
$W^{\pm}$ and $Z^0$ bosons are almost
indistinguishable in order of magnitude. It is also
intriguing that, in the region around $\varkappa \sim
1$, the presence or the absence of the
initial-gauge-boson electric charge is immaterial
because the change in the decay width is due primarily
to the increase in the phase space of final particles.

In the region of superstrong fields ($\varkappa \ge
10$), a sizable contribution to the total decay width
of the $Z^0$-boson comes from the $t$-quark-production
process $Z^0\rightarrow \bar t t$, which is forbidden
in a vacuum by conservation laws. As the
external-field strength increases, this reaction
becomes dominant in relation to other modes of
$Z^0$-boson decay, this being due to a very large mass
of the $t$-quark. The external field serves as some
kind of a catalyst for super heavy-particle-production
processes, which cannot occur under ordinary
conditions. In view of the aforesaid, it would be of
interest to discuss searches for new
(as-yet-undiscovered) hypothetical particles
(supersymmetry and so on) among new $Z^0$-boson decay
modes that arise in an external electromagnetic field.
Although external-field strengths necessary for
directly observing such reactions are not yet
available in experiments (see, for example,
\cite{Kurilin-1999}), these investigations are of
course very interesting from the point of view of
fundamental principles of physical theory.

\newpage
\begin{flushright}
\section*{APPENDIX}
\end{flushright}

In the present study, we have employed the special
mathematical functions $\Phi(z), \Phi'(z)$ and
$\Phi_1(z)$ generically referred to as Airy functions.
The function $\Phi(z)$ is a particular solution to a
second-order linear differential equation for specific
initial conditions:
$$
\Phi''(z)- z \Phi(z)=0, \quad
\Phi(0)=\frac{\pi}{3^{2/3}\Gamma(2/3)}, \quad
\Phi'(0)=-\frac{\pi}{3^{1/3}\Gamma(1/3)}.
 \eqno \mbox{(A.1)}
$$ It has the well-known integral representation $$
\Phi(z)=\int\limits_0^\infty \cos\left(zt+{t^3\over
3}\right)dt. \eqno \mbox{(A.2)}
$$
Two other functions
$\Phi'(z)$ and $\Phi_1(z)$ can be expressed in terms
of this function as
$$
 \Phi_1(z)=\int\limits_z^\infty
\Phi(t) dt , \qquad \Phi'(z)=\frac{d\Phi(z)}{dz} .
\eqno \mbox{(A.3)}
$$
 At small values of the argument
$z$ ($z\ll 1$), the Airy function $\Phi(z)$ can be
calculated on the basis of the following expansion in
a numerical series: $$ \Phi(z)={1\over
3^{2/3}}\sum_{n=0}^\infty \Gamma \biggl({n+1\over
3}\biggr) \sin \biggl( {2\pi\over 3}+{2\pi n\over 3}
\biggr) \frac{3^{n/3}}{n !}\cdot z^n. \eqno
\mbox{(A.4)}
$$
At large values of the argument ($z\gg
1$), the Airy function can be evaluated by using the
asymptotic expressions
$$ \Phi(z)={1 \over 2z^{1/4}
}\exp\left(-{2\over 3}z^{3/2}\right) \sum_{n=0}^\infty
(-1)^n \frac{\Gamma (3n+1/2)}{(2n)!\ 9^n}
\cdot\frac{1}{z^{3n/2}}, \eqno \mbox{(A.5)} $$

$$ \Phi(-z)={1\over z^{1/4}}\sum_{n=0}^\infty
\sin\left({2\over 3}z^{3/2}+{\pi\over 4}-\frac{\pi
n}{2}\right) \frac{\Gamma (3n+1/2)}{(2n)!\ 9^n}
\cdot\frac{1}{z^{3n/2}}. \eqno \mbox{(A.6)} $$ The
properties of the Airy functions are well known and
can be found in mathematical handbooks (see, for
example, \cite{Abramovitz}).


\end{document}